# HotOS XIX Panel Report: Panel on Future of Reproduction and Replication of Systems Research

*Roberta De Viti*, rdeviti@mpi-sws.org, Max Planck Institute for Software Systems
*Solal Pirelli*, solal.pirelli@epfl.ch, EPFL
*Vaastav Anand*, vaastav@mpi-sws.org, Max Planck Institute for Software Systems

At this year's HotOS, we organized and hosted a panel on the future of reproducibility and replication in the systems community. The purpose was twofold: to better understand the burning issues affecting the systems community and to improve the process of Artifact Evaluation (AE) at future systems conferences.

The panel consisted of three 20-minute open-mic discussion topics featuring panelists Shriram Krishnamurthi (Brown), Margo Seltzer (UBC), and Neeraja J. Yadwadkar (UT Austin). The three topics were:
- What is an artifact and how do we evaluate it?
- Is artifact evaluation leading to the outcomes we want?
- Do industry papers belong in industrial tracks or keynotes?

Each topic began with a brief opening statement from one of the panelists, followed by an opportunity for the audience to participate with questions, anecdotes, and additional insights.

In the remainder of this report, **we do not state personal opinions**, but focus on **summarizing the opinions of the participants**, pointing out where they have reached consensus. The discussion was conducted in a respectful manner, with active participation from all the audience. Throughout the panel, the participants converged on the following:

1. Artifacts for a paper should not be thought only as *software*, as it seems to be the case in practice; this fails to capture the full range of work undertaken by the systems community.

2. Artifact *availability* is paramount. The primary goal of artifact evaluation should be to ensure that the artifact is available to and reusable by the community; its availability can facilitate reproducibility and replication studies while enabling researchers to reuse and build upon it. In contrast, the primary goal should not be to verify if the claims made by authors are *reproducible*, as this reproducibility verification requires too much time investment for both authors and evaluators in the short timeframe of an AE.

3. There is the need to decouple *evaluating* an artifact and *reproducing or replicating* results. Although both are desired, the participants believe that the current AE process falls short of fully supporting either objective. Most audience members deemed the quality of reproduction or replication efforts well below the required standards; furthermore, they felt the current AE process is primarily designed for assessing *software* artifacts, which frequently marks research requiring specialized or custom hardware solutions as not *re*producible.

In the rest of this report, we provide more details on the discussions that took place, and summarize other noteworthy points raised during the panel.

# 1. What is an artifact?

*"Not all artifacts are code"*

Most members of the audience believe that the systems community frequently makes an incorrect assumption in Artifact Evaluation procedures: that all artifacts should be treated as software and can be conveniently packaged into a virtual machine or container. The current AE procedure has posed ongoing challenges for community members working on hardware-related projects, or software that necessitates specific hardware components or dependencies that do not align with the current AE guidelines. Furthermore, even software systems cannot easily be reduced to a piece of software that can be packaged on one machine and easily rebuilt, reconfigured, and re-run to obtain the same results on a different machine – at least, not without extensive efforts from both sides, authors and evaluators.

The audience shared various anecdotes that reinforced this claim, one of which was the case of Enzian: a research computer developed by the ETH systems group. The main contribution of the Enzian paper (ASPLOS'22) lies in its novel hardware. Consequently, it has been highly challenging for artifact evaluators to obtain physical access to the system, let alone reproduce the specific piece of hardware. While the AE process for ASPLOS was eventually adjusted to facilitate result reproduction for Enzian via remote access, [this was a lot of work](#) for both authors and evaluators!

The audience kept bringing up other examples, despite falling under the current definitions of an artifact *in theory*, they may not be well supported in the procedures for giving out badges. For instance, software that requires proprietary hardware, software that necessitates a long-running process to reproduce results, or execution traces (e.g., [Serverless in the Wild: Characterizing and Optimizing the Serverless Workload at a Large Cloud Provider](#), [Characterizing Microservice Dependency and Performance: Alibaba Trace Analysis](#)).

In summary, the HotOS community thinks that, in practice, the AE guidelines do not effectively apply to all artifacts built by the systems community today, but rather focus on high-level software artifacts… to the point that some community members are afraid that the current AE process might dissuade students from working on research topics where producing artifacts is more difficult!

# 2. Reproducibility! But at what COST?

The discussion about the Enzian project led to another, related issue: AE is very time consuming, but is this effort worth it? Let's take a step back and look at the big picture.

**The evaluators.** Currently, the AE evaluation process mainly involves early-career researchers, such as junior professors and students. Initially, this approach was adopted to address the issue of "sub-reviewer" abuse, where faculty members would delegate reviewing tasks to their students but take credit for the reviews themselves. By directly involving students in the AE process, we could ensure they receive the credit they deserve. However, it appears that this solution is not as effective in practice: students are spending significant amounts of time on AE, and they are not benefiting significantly in terms of learning from the process.

Another concern is that some of these evaluators of the artifacts lack sufficient experience. First- or second-year students may not possess the expertise to thoroughly test for reproducibility and replication, although they can assess availability and functionality. Additionally, there seems to be a lack of accountability for evaluators, as those who perform poorly in the artifact evaluation face no consequences.

**The authors.** On the other side of the AE process, there are the authors, who are mainly other students. There is a significant burden on them to produce "production-ready" artifacts. They might spend months meticulously preparing the artifacts according to the requirements, and yet again, this effort often results in minimal learning for them.

In summary, the real concern seems to be time spent on both sides to *reproduce* results. But why are we reproducing results? Is that the right way to evaluate an artifact? That's what the HotOS community kept debating.

# 3. How should we evaluate an artifact?

Currently, the AE process checks for (optional) availability, functionality, and reproducibility. However, there seem to be two issues:

**How the badges were chosen.**
Most of the attendees kept wondering: whom did ACM consult before agreeing upon badges? No one in the HotOS audience seemed to know the answer! Furthermore, there seems to be a mismatch between the definitions of various terms used in AE (e.g., the definition of reproduction that the ACM badge uses differs from the definitions in the [Feitelson](Feitelson) paper).

**The outcome of having three badges.**
The community fears that the lack of badges might cause some papers to be dismissed. Would a paper that does not have all the badges be considered less rigorous? Was this outcome intended? Would this outcome be good? There didn't seem to be consensus on the answers. Furthermore, as mentioned above, some participants fear that the current badge assignment procedure could potentially bring an undesired effect: discouraging students from doing work that cannot get all three badges. But why are there three badges? What are the goals of artifact evaluation?

**Revising the goals of artifact evaluation.**
Is the goal to ensure that all the claims of the paper are verified by the artifact? Or is the goal to ensure that the artifact provides a way to verify the claims in the future? While both these goals have advantages, what should AE focus on, specifically?
The participants reached consensus on the following: the goal should be to check that the artifact provides what the paper promised – not just in terms of results but in terms of the overall claims made in the paper. This goal will have a positive side effect: to persuade authors to sharpen their claims and be explicit about the shortcomings of their system. For example, if a tool cannot handle Java programs with multithreading, or a C-analysis tool cannot handle recursive calls, then the authors would have to be explicit about this in their paper, to avoid a mismatch between the artifact and the claims in the paper.

**Dropping the reproducibility badge.**
Finally, most attendees wondered whether we should *drop reproducibility badges*. Checking for reproducibility takes a lot of time and may not be reasonably doable within the timeframe of an AE, and the audience felt that the benefits are not worth the cost. Thus, the discussion kept circling around the issue of *reproducibility*, and whether it should be a goal for AE at all…

A key question asked by a member of the audience was the following: reproducibility has been introduced in other research fields (e.g., psychology) because there has been evidence of fraud in the area, which had to be contained. However, is this actually a concern in CS? Is there evidence of fraud in our field that demands to reproduce the results of every paper? Most people seem to agree that *no*, there is not enough evidence of fraud that would justify a thorough check for reproducibility, and "there is no reason for the AECs to be the police". At most, there might be sloppiness, which calls for a sanity check. However, such sanity check is currently demanding a very high cost; for this reason, the participants seem to agree on the fact that reproducibility is better served to be left as an exercise for future work, and a paper artifact should not be treated as an undergraduate assignment in need of grading.

**Focusing on a broader definition of availability.**
In conclusion, most attendees agreed that *availability* is the key for artifacts, and what retains most value. However, the current definition used by "available" badges felt too restricted for many attendees, for instance because it does not even require long-term storage, or that the artifact be actually usable. To better test for availability and functionality, there should be other forms of evidence allowed: e.g., video evidence, the possibility for remote access. These forms of evidence should suffice to show to AECs that certain artifacts are available and functional.

## 4. Industry papers

*"Industry papers are real systems that should not require reproducibility"*

According to that subset of participants that work in companies, it is not possible, in practice, to have industry papers release artifacts. First, industry papers cannot release neither artifacts nor raw results for legal (as well as competitive) reasons. Second, industry software is held at a higher standard than "grad-student code", which means that releasing code would be far more time consuming.

However, even though industry papers might not be reproducible, they usually refer to systems that *have been deployed* and are effectively widely used. According to some of the participants, this implies that the results are indeed *real* ("unless we are dealing with a company like Theranos"). Besides, the key contribution of industry papers is the problems and *the ideas\** rather than the numbers being reported. Thus, it is considered not important for industry papers to be reproducible. More broadly, the goal of systems research in general is not to produce artifacts, rather to produce and communicate ideas. In the case of industry papers, we have to make the compromise that we cannot have both, as we are trying to do in academic research: we cannot have artifact availability, but we can have access to their problems and ideas. If we force availability making it mandatory to submit to conferences, we will miss out on the ideas.